\def\cm{{\rm\thinspace cm}}

\def\erg{{\rm\thinspace erg}}

\def\keV{{\rm\thinspace keV}}

\def\s{{\rm\thinspace s}}

\def\ergpcmsqps{\hbox{$\erg\cm^{-2}\s^{-1}\,$}}

\def\ergps{\hbox{$\erg\s^{-1}\,$}}

\def\psqcm{\hbox{$\cm^{-2}\,$}}

\def\gs{\mathrel{\raise0.35ex\hbox{$\scriptstyle >$}\kern-0.6em
\lower0.40ex\hbox{{$\scriptstyle \sim$}}}}
\def\ls{\mathrel{\raise0.35ex\hbox{$\scriptstyle <$}\kern-0.6em
\lower0.40ex\hbox{{$\scriptstyle \sim$}}}}
\documentstyle[psfig,times]{mn}
\begin{document}

\title[{\it Chandra}--SCUBA sources]
{Testing the connection between the X-ray and submillimetre source
populations using {\it Chandra}}
\author[Fabian et al.]
{A.C.\ Fabian$^1$, Ian Smail$^2$, K.\ Iwasawa$^1$, S.W.\ Allen$^1$,
A.W.\ Blain$^1$, C.S.\ Crawford$^1$, \and S.\ Ettori,$^1$ R.J.\
Ivison$^3$, R.M.\ Johnstone$^1$, J.-P.\ Kneib$^4$ and R.J.\
Wilman$^1$\\
\vspace*{1mm}\\
$^1$ Institute of Astronomy, Madingley Road, Cambridge CB3 0HA \\
$^2$ Department of Physics, University of Durham, South Road, Durham,
DH1 3LE\\
$^3$ Department of Physics \& Astronomy, University College London, Gower
Street, London, WC1E 6BT\\
$^4$ Observatoire Midi-Pyr\'en\'ees, 14 Avenue E.\ Belin, 31400 Toulouse, 
France
}
\maketitle

\begin{abstract}
The powerful combination of the {\it Chandra} X-ray telescope, the
SCUBA submillimetre-wave camera and the gravitational lensing effect
of the massive galaxy clusters A\,2390 and A\,1835 has been used to
place stringent X-ray flux limits on six faint submillimetre SCUBA
sources and deep submillimetre limits on three {\it Chandra} sources
which lie in fields common to both instruments. One further source is
marginally detected in both the X-ray and submillimetre bands. For all
the SCUBA sources our results are consistent with starburst-dominated
emission. For two objects, including SMMJ\,14011+0252 at $z=2.55$, the
constraints are strong enough that they can only host powerful active
galactic nuclei if they are both Compton-thick and any scattered X-ray
flux is weak or itself absorbed. The lensing amplification for the
sources is in the range 1.5--7, assuming that they lie at $z\gs 1$.
The brightest detected X-ray source has a faint extended optical
counterpart ($I\approx22$) with colours consistent with a galaxy at
$z\simeq 1$. The X-ray spectrum of this galaxy is hard, implying
strong intrinsic absorption with a column density of about
$10^{23}\psqcm$ and an intrinsic (unabsorbed) 2--10~keV luminosity of
$3\times 10^{44}\ergps$. This source is therefore a Type-II quasar.
The weakest detected X-ray sources are not detected in {\it HST}
images down to $I\simeq 26$.
\end{abstract}


\section{Introduction}
The spectrum of the X-ray Background (XRB) over the 1-7~keV band is a
power-law of energy index 0.4 (Gendreau et al.\ 1995) which is flatter
than that of any known class of extragalactic source. Following the
original suggestion of Setti \& Woltjer (1989), it is commonly assumed
that this is due to the XRB being dominated by many absorbed sources.
These sources, with different absorbing column densities and
redshifts, combine to give the observed XRB spectrum (Madau,
Ghisellini \& Fabian 1994; Matt \& Fabian 1994; Comastri et al.\ 1995;
Wilman \& Fabian 1999). The absorbed energy is presumably reradiated
in the Far InfraRed (FIR) band, as observed in nearby heavily absorbed
Active Galactic Nuclei (AGN) like that in the luminous {\it IRAS}
galaxy NGC\,6240 (Vignati et al.\ 1999). The residual harder X-ray
emission which penetrates the absorbing medium, and the tail of the
reradiated emission in the submillimetre band from such an source both
exhibit a negative K-correction. This means that obscured AGN should
be detectable to large redshifts in both bands.

Observations of the XRB with {\it Chandra} show that much (more than
80 per cent) of it is resolved into point sources in the 2--8~keV band
(Brandt et al.\ 2000; Mushotzky et al.\ 2000). More than half the
intensity is due to a combination of hard sources identified with
either otherwise normal bright galaxies or in optically faint or even
invisible galaxies.  Deep 850-$\mu$m submillimetre observations with
SCUBA show that much (more than 80 per cent) of the submillimetre
background seen by {\it COBE}-FIRAS (Fixsen et al.\ 1998) at this
wavelength is resolved into discrete sources (Blain et al.\ 1999).

A key question is what fraction of the deep X-ray and submillimetre
source populations are related. The far-infrared and submillimetre
background represents a significant part of the energy output of
objects in the Universe. If the contribution of starbursts and AGN to
the background can be separated using deep X-ray images, then the
relative importance of high-mass star formation and AGN in heating the
dust responsible for this emission can be determined.  Recent
modelling suggests that the AGN fraction in the submillimetre background is
about 20 per cent (Almaini, Lawrence \& Boyle 1999; Fabian \& Iwasawa
1999; Gunn \& Shanks 1999). The uncertainty is such that AGN may
contribute in total between 10--50 per cent of the total energy output
of stars (Fabian 1999).

Smail et al.\ (1997, 1998) have used the SCUBA instrument at the JCMT
(Holland et al 1999) to study the submillimetre sources in seven
massive clusters of galaxies at redshifts between 0.2 and 0.4,
exploiting the gravitational lensing magnification from the clusters to
make an exceptionally deep 850-$\mu$m survey for background sources.
{\it Chandra} images of two of the clusters, A\,2390 and A\,1835, have
recently been obtained which, owing to the superb angular resolution,
probe much deeper than any previous X-ray images of these regions, for
example the {\it ROSAT}-HRI limit of $8 \times
10^{-14}$\,erg\s$^{-1}$\,cm$^{-2}$ to the 0.1--2.0\,keV X-ray flux of
the SCUBA source SMMJ\,14011+0252 in A\,1835 (Ivison et al.\ 2000).
Here we study the {\it Chandra} X-ray limits to the flux of the SCUBA
sources and conversely SCUBA limits on the {\it Chandra} sources
found in the images. Only one source is marginally detected in both
wavebands. We then discuss the implications for obscured AGN models.

\section{Observations and Results}

{\it Chandra} observed A\,2390 on 1999 November 5 for a livetime of
9,126~s and A\,1835 on 1999 December 11 for a total of 19,626~s. For
each observation the cluster lies close to the aimpoint of the
telescope on the ACIS-S back-illuminated CCD chip. Only mild temporal
variations in count rate are seen through the observations so we use
the whole exposure in this work. A 7-arcsec pointing offset evident in
the A\,1835 field from the position of the cluster X-ray peak and
several other source identified in the field has been removed. We estimate that
the source positions from the {\it Chandra} images are accurate
to 1$''$ rms.

The analysis of the cluster emission will be presented elsewhere;
here we restrict ourselves to the positions of the SCUBA sources and
of three X-ray sources in the {\it Chandra} fields which lie within the
SCUBA field (i.e.\ within 1.5~arcmin of the cluster centre; Table
1). A further possible ($2.8\sigma$) X-ray source in the A\,2390 field
is also discussed since it is spatially coincident with a tentative
($2\sigma$) SCUBA source. No point-like X-ray sources are seen in the
SCUBA region of the {\it Chandra} field of A\,1835. 

The X-ray data were analysed using images in the 0.5--2~keV and
2--7~keV bands with one-arcsec pixels. The sources detected in the
SCUBA field of A\,2390 appear principally in just 4 neighbouring
Chandra pixels (a few counts from the brightest sources occupy 2
further pixels). We therefore adopt a region of 4 sq.\ arcsec when
estimating source fluxes and limits. The cluster emission means that
the background is not flat, and so the background has been determined
by averaging the 8 neighbouring pixels in images formed with 10 by 10
arcsec pixels, excluding the source pixel.  Upper limits in the X-ray
band have been obtained by using the Bayesian method of Kraft, Burrows
\& Nousek (1990) and are quoted at the 99 per cent confidence level
(Table 2).  We assume Galactic columns of $6.8\times 10^{20}\psqcm$
for A\,2390 and $2.3\times 10^{20}\psqcm$ for A\,1835.

The submillimetre, radio and optical imaging of A\,1835 used here is
discussed in detail in Ivison et al.\ (2000), while the submillimetre
observations of A\,2390 are detailed in Smail et al.\ (1998; a new
sources is reported here), the VLA 1.4-GHz radio map in Edge et al.\
(1999) and the optical imaging of this cluster with {\it Hubble Space
Telescope} (HST) is described in Pell\'o et al.\ (1999).  In addition,
there is deep 6.7$\mu$m and 15-$\mu$m {\it ISO} imaging of A\,2390
(Altieri et al.\ 1999; Lemonon et al.\ 1999), which is sensitive to
hot dust emission from background galaxies out to $z\sim 1$.

\begin{figure*}
\centerline{\psfig{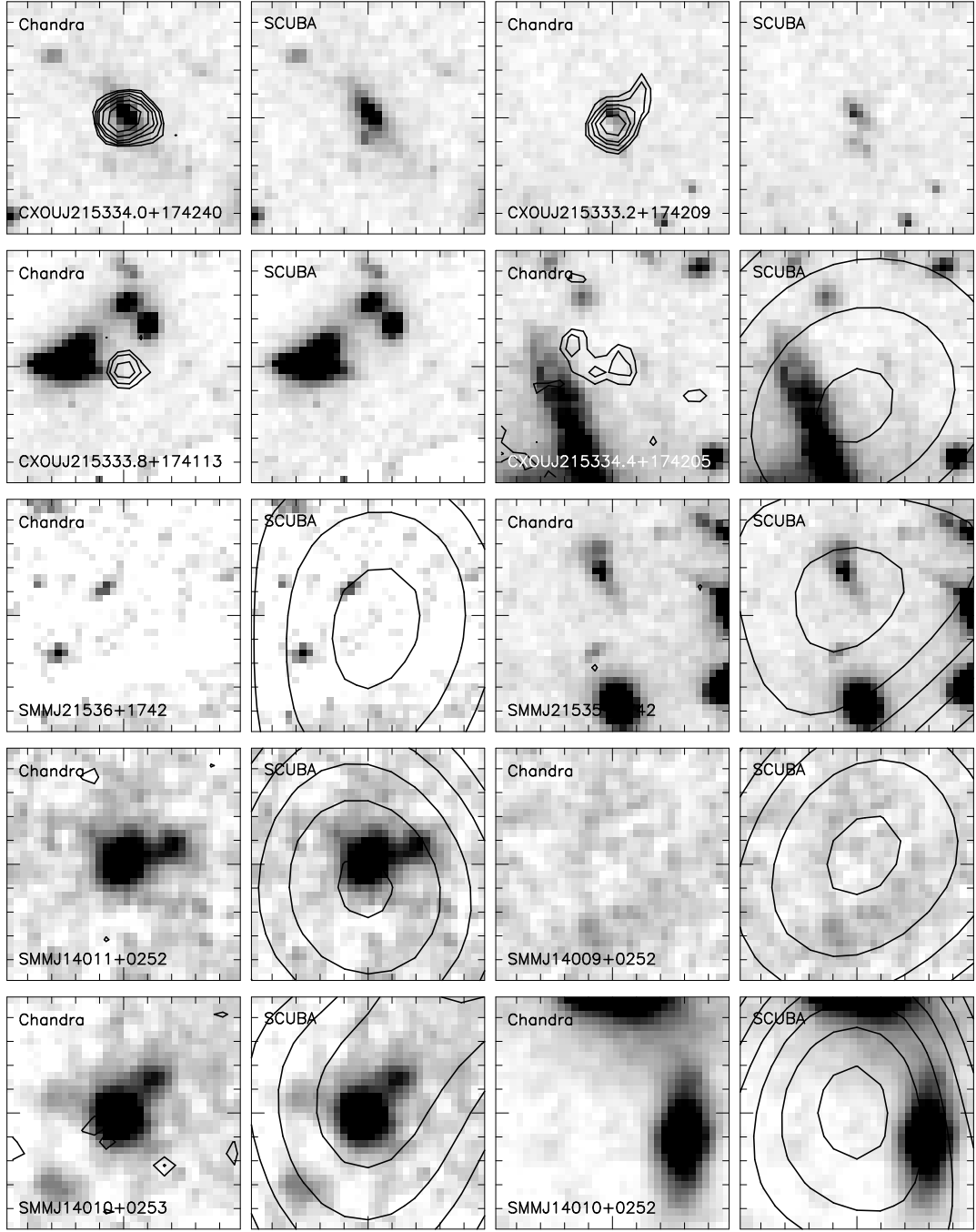}}
\caption{
Optical, X-ray and submm views of the ten {\it Chandra} and SCUBA
sources lying within the SCUBA maps of A\,2390 and A\,1835.  For each
source we show two panels, in the left-hand panel we overlay the {\it
Chandra} image on the optical images of the field, the right-hand
panel shows the equivalent view with the SCUBA map overlayed.  The
optical image for A\,1835 is the ground-based $I$-band exposure from
Ivison et al.\ (2000), while for A\,2390 we have combined the {\it
HST} WFPC2 F555W and F814W images (Pell\'o et al.\ 1999) and rebinned
these to the scale of the A\,1835 $I$-band image to enhance the
visibility of faint extended features.  Each panel is 10 arcsec
square, with north top and east left.  The {\it Chandra} images have
been convolved with a 0.8$''$-FWHM gaussian for display purposes.}
\end{figure*}

In Fig.~1 we show overlays of the {\it Chandra} and SCUBA images on
optical images of A\,1835 and A\,2390.  We note that the brightest {\it
Chandra} source, CXOUJ215334.0+174240, is associated with a relatively
large (4$''$ total extent) mid-type spiral with a blue nucleus and a
faint companion.  This galaxy has a colour and apparent magnitude
similar to those expected for a slightly reddened $L^\ast$ mid-type spiral
at $z=0.85\pm0.15$.  While the next brightest X-ray source,
CXOUJ215333.2+174209 has a more amorphous counterpart, with a bright
compact nucleus and an asymmetric envelope and has a much bluer colour,
its redshift is not strongly constrained by the current data.
Counterparts to the remaining two {\it Chandra} sources are not seen in
the optical imaging (although there is a very faint object in the
vicinity of CXOUJ215334.4+174205); if they are background galaxies
then, after correction for lensing amplification, they must be fainter
than $I\sim 27$.  CXOUJ215334.0+174240 is the only one of the four {\it
Chandra} sources detected at 1.4~GHz with the VLA, at a flux density of
$2.6\pm 0.1$ mJy, the 3-$\sigma$ limits on the remainder being $<0.2$
mJy.

For the two X-ray sources which are detected in both X-ray bands we
have determined an X-ray spectral index, $\alpha_{\rm x}$ from the
flux ratios. These are both unphysically negative ($-0.8$ and $-0.3$;
we define all spectral indices according to $F\propto \nu^{-\alpha}$)
and thus indicate intrinsic absorption. The brightest source,
CXOUJ215334.0+174240, has $\sim 90$ counts, which enables a crude
spectral analysis to be performed. A straight power-law spectrum with
Galactic absorption is a poor fit and yields an energy index of
$-0.3$. Including additional soft X-ray absorption allows a better
fit, although with no firm constraints on the index. The absorption
however must exceed $N=6\times10^{21}\psqcm$ at the 90 per cent
confidence level. Note that the intrinsic absorption will be
approximately $(1+z)^3 N$, where $z$ is the redshift of the source.
For an assumed energy index of 1, typical of quasars, and the redshift
indicated by the optical colours, we find
$N=(6\pm2)$--$(9\pm3)\times10^{22}\psqcm$, and an unabsorbed intrinsic
2--10~keV luminosity of 2.7--$6.8\times 10^{44}\ergps$, for
$z=0.7$--1, respectively. After correction for the lensing
amplification factor of about 2, the source has an intrinsic
$L(2$--$10\keV)\approx 2$--$3\times10^{44}\ergps$. This makes it a
strong contender to be one of the first genuine Type-II quasars (see
discussions in Halpern et al.\ 1999; Vignati et al.\ 1999;
Franceschini et al 1999).

The X-ray limits on the SCUBA sources in the A\,1835 field are about
100 times deeper than previously achieved using {\it ROSAT} data
(Ivison et al.\ 2000). We have produced two submillimetre-to-X-ray
spectral indices, $\alpha_{SX}$, using the measured values at
850$\mu$m and 2~keV in our rest frame. Rather than obtain one X-ray
estimate for the spectral flux at 2~keV from a whole band measurement,
which would be biased by the most sensitive lower energy end of the
band, we have estimated it by converting the observed counts to fluxes
in the 0.5--2 and 2--7~keV bands (listed in Table 2), using the
response matrix of the ACIS-S chip and assuming an intrinsic energy
index of 1 (which is roughly appropriate for the scattered flux from
an absorbed AGN).

\begin{table*}
\begin{center}
\caption{Postions of the {\it Chandra} (prefixed CXOU) and SCUBA
sources (prefixed SMM) discussed here. We list the $1''$-diameter
photometry of the {\it Chandra} sources from the {\it HST} F555W and
F814W frames and for all sources the gravitational lensing
amplification obtained by modelling the mass distribution of the
clusters (see Blain et al.\ 1999 for details: * indicates no solution
at that redshift). Known spectroscopic redshifts are put in the left
column and the amplification at that redshift in the right.}

\begin{tabular}{lllcccc}
Source Name & R.A. & Dec & $V_{555}$  & $I_{814}$ & \multicolumn{2}{c}{Amplification}\\
& \multicolumn{2}{c}{J2000} & & & $z=1$ & $z=2.5$ \\
CXOUJ215334.0+174240 & 21 53 34.0  & 17 42 40   & $25.6\pm 0.1$ &
$22.6\pm 0.0$ & $z\sim 1$ & 1.9 \\
CXOUJ215333.2+174209 & 21 52 33.2  & 17 42 09   & $25.7\pm 0.1$ & $24.2\pm 0.1$ & 3.9 & 6.9 \\
CXOUJ215333.8+174113 & 21 53 33.76 & 17 41 13   & $>26.2$ & $>24.5$ & 7.0 &  * \\
CXOUJ215334.4+174205 & 21 53 34.4  & 17 42 05   & $>26.9$ & $>26.2$ & 1.7 & 1.9 \\
SMMJ\,21536+1742     & 21~53~38.5  & 17~42~19   & ... & ... & 2.1 & 2.6 \\
SMMJ\,21535+1742     & 21~53~33.2  & 17~42~49   & ... & ... & 1.7 & 1.9 \\
SMMJ\,14011+0252     & 14~01~04.96 & 02~52~23.5 & ... & ... & $z=2.55$ & $3.0\pm0.6$ \\
SMMJ\,14009+0252     & 14~00~57.55 & 02~52~48.6 & ... & ... & 1.4 & 1.5 \\
SMMJ\,14010+0253     & 14~01~03.09 & 02~53~12.0 & ... & ... & $z=2.22$ & $4.8\pm2.8$ \\
SMMJ\,14010+0252     & 14~01~00.53 & 02~51~49.4 & ... & ... & 1.6 & 1.8 \\
\end{tabular}         
\end{center}
\end{table*}
 
\begin{table*}
\begin{center}
\caption{Count rates and fluxes of the {\it Chandra} and SCUBA sources. The total 
{\it Chandra} background-subtracted count for each source, the background 
(per square arcsec), and the flux, obtained assuming an X-ray energy 
index of 1 are shown in two X-ray bands:  0.5--2~keV (no brackets) and 
2--7~keV (brackets). The final column give the submillimetre-to-X-ray index 
$\alpha_{\rm SX}$ obtained in the same bands (with the same bracket convention).}

\begin{tabular}{lccccc}
Source Name & \multicolumn{3}{c}{X-ray flux -- at low (high) energy} & 
{SCUBA 850-$\mu$m} & {$\alpha_{SX}$} \\
& Count & Background & Flux & Flux density & \\ 
& & & ($10^{-15}\ergpcmsqps$) & (mJy)  & (99\% c.l.) \\
[5pt]
CXOUJ215334.0+174240 & $32\pm5.8$ ($54.1\pm7.3$) & 0.34 (0.15) & $9.9$
($90$) & 
$<5.5$ & $<1.06$ ($<0.91$) \\
CXOUJ215333.2+174209 & $19\pm4.6$ ($14.3\pm3.9$) & 0.5 (0.2) & 5.9 (23) & 
$<5.7$ & $<1.10$ ($<1.01$) \\
CXOUJ215333.8+174113 & $11.9\pm3.6$ ($<13$) & 0.28 (0.13) & 3.7 ($<17$) & 
$<5.7$ & $<1.14$ ($<1.03$) \\
CXOUJ215334.4+174205 & $10.4\pm3.7$ ($<7.9$) & 0.89 (0.18) & 3.2 ($<13$) & 
$3.9\pm2.2$ & 1.12 (1.02) \\
SMMJ\,21536+1742 & $<10.6$ ($<4.6$) & 1.1 (0.68) & $<3.2$ ($<7.7$) & 
$6.7\pm1.2$ & $>1.16$ ($>1.10$) \\
SMMJ\,21535+1742 & $<6.0$ ($<7.9$) & 2.4 (0.48) & $<1.8$ ($<13$) & 
$8.3\pm 2.2$ & $>1.21$ ($>1.07$) \\
SMMJ\,14011+0252 & $<5.6$ ($<5.6$)& 7.3 (2.4) & $<0.78$ ($<4.4$) & 
$14.6\pm 1.8$ & $>1.32$ ($>1.19$)  \\
SMMJ\,14009+0252 & $<6.0$ ($<4.6$) & 1.1 (0.6) & $<0.84$ ($<3.6$) & 
$15.6\pm 1.9$ & $>1.32$ ($>1.21$)  \\
SMMJ\,14010+0253 & $<6.2$ ($<4.6$) & 7.6 (2.2) & $<0.87$ ($<0.87$) & 
$4.3\pm 1.7$ & $>1.23$ ($>1.12$)  \\
SMMJ\,14010+0252 & $<6.7$ ($<6.0)$ & 2.4 (1.0) & $<0.94$ ($<4.7$) & 
$4.2\pm 1.7$ & $>1.22$ ($>1.10$) \\
\end{tabular}
\end{center}
\end{table*}

\section{Discussion}

We compare expected values of $\alpha_{\rm SX}$ as a function of
redshift in Fig.~2 for various classes of observed objects, both
starbursts and AGN. We select 3C\,273 as an example of a powerful
quasar, and use the submillimetre data of Neugebauer, Soifer \& Miley
(1985) in our model; Arp\,220 as a starburst using the submillimetre
data of Sanders et al.\ (1999) and the X-ray data of Iwasawa (1999);
and NGC\,6240 as a powerful obscured (Compton-thick) AGN using an
estimate of the submillimetre spectrum based on that from Arp\,220
normalised by the FIR luminosity and the {\it BeppoSAX} spectrum from
Vignati et al.\ (1999). We also show how $\alpha_{\rm SX}$ changes as
the absorption in NGC\,6240 decreases from the measured value of
$2\times 10^{24}$ to $5\times 10^{23}\psqcm$ and also if the scattered
fraction drops to 1 per cent.  Data from several quasars at $z>4$ are
shown for comparison with the 3C\,273 prediction.

\begin{figure}
\centerline{\psfig{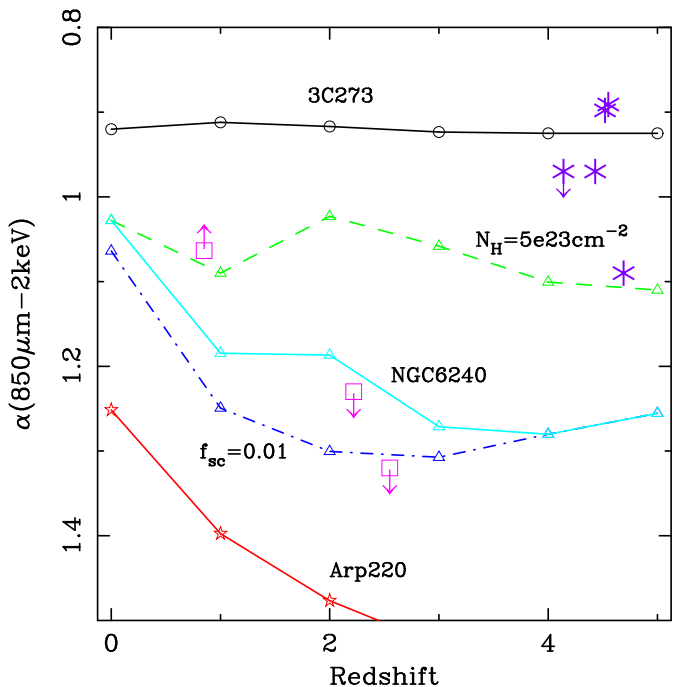}}
\caption{Submillimetre to X-ray spectral index, $\alpha_{\rm SX}$,
plotted against redshift for a powerful quasar, 3C\,273, an absorbed
AGN, NGC\,6240, and a starburst, Arp\,220. The effects of decreasing
the scattered fraction from NGC\,6240 to 1 per cent, and of its column
density to $5\times 10^{23}\psqcm$ are indicated by the dash-dot and
dash lines. The {\it Chandra}/SCUBA sources with known redshifts are
indicated by the squares. The $\ast$ symbols represent the five $z>4$
quasars for which both X-ray and SCUBA (850$mu$m) data are
available; monochromatic X-ray fluxes at 2 keV were calculated from
the (Galactic) absorption-corrected 0.1--2.0 keV fluxes in Kaspi et
al.~(2000) assuming a power-law of $\Gamma=2$; 850-$\mu$m fluxes are
taken from the compilation in McMahon et al.~(1999).}
\end{figure}

The typical limits on $\alpha_{\rm SX}$ for the SCUBA galaxies are
$\alpha_{\rm SX} \ls 1.2$ (at 99 per cent confidence).  Combining the
limits for the 6 galaxies we obtain a 99 per cent confidence limit on a
typical SCUBA galaxy of $\alpha_{\rm SX}> 1.34$, with the strongest
constraints for individual galaxies being $\alpha_{\rm SX} > 1.32$  for
SMMJ\,14011+0252 ($z=2.55$) and SMMJ\,14009+0252 in the A\,1835 field.
Thus the  indices for this population are straightforwardly consistent
with starbursts. They do not resemble Compton-thin AGN at any redshift
and are inconsistent with the spectrum of NGC\,6240 at any redshift.
If these galaxies host a powerful AGN then either a) reprocessed
radiation from the AGN provides only a minor fraction of the
submillimetre luminosity, or b) the source must be Compton thick and in
addition the fraction of any scattered X-ray emission must be less than
one per cent.  We note that X-ray limits on the hyperluminous {\it
IRAS} galaxy F15307+3252 at $z=0.92$, which does contain an AGN since
broad scattered emission lines are seen (Hines et al.\ 1995), show that
it must have a very low X-ray scattered fraction ($<1$ per cent),
possibly due to intrinsic absorption of the scattered emission (Fabian
et al.\ 1996). If this is typical of obscured AGN more powerful than
NGC\,6240, then the present results would be consistent with obscured
AGN provided that they are at relatively high redshift ($z\gs 2$).

Of the four {\it Chandra} sources we identified in our fields, two have
probable optical counterparts suggesting that they are distant disk
galaxies.  The colours and luminosity of the brightest of these suggest
it is an $L^\ast$ mid-type galaxy at $z\sim 1$.  If the two
optically-unidentified {\it Chandra} sources are AGN, and the host
galaxies have luminosities of $L^\ast$ or greater, then they must be at
$z>2$, or be intrinsically reddened. Obscuration of both AGN and
surrounding spheroid is required by some models for the XRB (Fabian
1999).  It is possible for these dust-enshrouded strongly-obscured AGN
to be undetected in our 850-$\mu$m SCUBA maps  if their dust is
typically much hotter than 40\,K.  Comparing the radio flux and SCUBA
limit for the brightest {\it Chandra} galaxy CXOUJ215334.0+174240
(\S2), the redshift $z$ and dust temperature $T_d$ of the source must
satisfy the relationship $T_d > 27(1+z)$\,K.  Taking the redshift
constraint from the optical, $z \simeq 0.85\pm 0.15$, we derive $T \ge
50$\,K.  Such sources would not contribute substantially to the
submillimetre background at $\sim 1$\,mm, but might at shorter
wavelengths.  Eventual comparison with  {\it ISO} imaging may shed
further light on the nature of the dust emission in these sources.

In summary, {\it Chandra} and SCUBA observations have been combined to
probe the relation between the faint X-ray and submillimetre sky. Only
one marginal source is seen in both datasets, and so in general we
find deep submillimetre limits from SCUBA on the X-ray sources found
using {\it Chandra} and deep X-ray limits from {\it Chandra} on the
SCUBA-selected sources. The limits on background sources in these
fields are particularly strong due to amplification by gravitational
lensing.

For the SCUBA galaxies, we cannot completely rule out the presence of
either an obscured, weak AGN or a more powerful Compton-thick AGN in
which any scattered flux is also  very weak or absorbed.  However, the
simplest explanation of our results on the SCUBA sources is that they
are predominantly powered by starbursts.  Clearly the current sample is
small and so we cannot make any definitive statement about the whole
SCUBA population. We note that at the moment our conclusions are not
inconsistent with suggestions that AGN powering 20 per cent of the
SCUBA population, although a substantially higher fraction would be
difficult to accommodate.  

For the {\it Chandra} sources in the A2390 field we find that they
have optically faint counterparts, $I\gs 23$--27. We
identify one source as a probable Type-II obscured quasar at $z\simeq
1$.  We suggest that the remaining, typically fainter sources are
either more distant, $z>2$, or intrinsically reddened.  Observations
of these fields in the near-infrared are urgently needed to test the
nature of these sources.

\section{Acknowledgements}
We thank L.\ van Speybroeck and his colleagues for the superb X-ray
telescope and M.\ Weisskopf and the project team for the {\it Chandra}
mission. ACF is grateful to NASA for the opportunity to participate in
{\it Chandra}. The Royal Society is thanked for support by ACF, IRS, SWA,
CSC and SE   and PPARC by RJI. AWB and JPK thank the Raymond \&
Beverly Sackler Foundation and the LENSNET European Network,
respectively.

\end{document}